\documentclass[a4paper,twoside,12pt]{article}
\input{obsjkt.sty}

\newcommand{\Msun}{\ensuremath{~{\rm M}_\odot}}                   
\newcommand{\Rsun}{\ensuremath{~{\rm R}_\odot}}                   
\newcommand{\rhosun}{\ensuremath{~\rho_\odot}}                    
\newcommand{\Teff}{\ensuremath{T_{\rm eff}}}                      
\newcommand{\Vsys}{\ensuremath{V_\gamma}}                         
\newcommand{\EBV}{\ensuremath{E(B\!-\!V)}}                        
\newcommand{\degr}{\ensuremath{^\circ}}                           
\renewcommand{\kms}{~km~s$^{-1}$}                                 
\newcommand{\mc}[1]{\multicolumn{2}{c}{#1}}
\newcommand{\corot}{\textit{CoRoT}}

\newcommand{\gaia}{\textit{Gaia}}
\newcommand{\targ}{CW~Eri}
\newcommand{\targfull}{CW Eridani}

\newcommand{\Msunnom}{\hbox{$\mathcal{M}^{\rm N}_\odot$}}
\newcommand{\Rsunnom}{\hbox{$\mathcal{R}^{\rm N}_\odot$}}
\newcommand{\Lsunnom}{\hbox{$\mathcal{L}^{\rm N}_\odot$}}

\usepackage{rotating}

\begin{document} 

\OBSheader{Rediscussion of eclipsing binaries: \targ}{S.\ Overall and J.\ Southworth}{2024 April}

\OBStitle{Rediscussion of eclipsing binaries. Paper XVII. \\ The F-type twin system CW Eridani}

\OBSauth{Stephen Overall and John Southworth}

\OBSinstone{Astrophysics Group, Keele University, Staffordshire, ST5 5BG, UK}

\OBSabstract{\targ\ is a detached eclipsing binary system of two F-type stars with an orbital period of 2.728~d. Light curves from two sectors of observations with the Transiting Exoplanet Survey Satellite (TESS) and previously published radial velocity data are analysed to determine the system's physical properties to high precision. We find the masses of the two stars to be $1.568 \pm 0.016 \Msun$ and $1.314 \pm 0.010 \Msun$, the radii to be $2.105 \pm 0.007\Rsun$ and $1.481 \pm 0.005\Rsun$ and the system's orbit to have an eccentricity of $0.0131 \pm 0.0007$. The quality of the TESS photometry allows the definition of a new high-precision orbital ephemeris, however no evidence of pulsation is found. We derive a distance to the system of $191.7\pm3.8 \mathrm{\,pc}$, a value consistent with the \gaia\ DR3 parallax which yields a distance of $187.9^{+0.6}_{-0.9}$ pc. The measured parameters of both stellar components are found to be in agreement with theoretical predictions for a solar chemical composition and an age of 1.7~Gyr.}


\section*{Introduction}

Detached eclipsing binaries (dEBs) are a vital source of stellar parameters as they allow direct measurement of the component stars' physical properties when combining light curves and radial velocity (RV) observations \cite{Andersen91aarv,Torres++10aarv,Me15aspc}. Detached systems are particularly useful as, in the absence of mass transfer, the components are representative of single stars and are therefore an invaluable source of data for testing and refining stellar evolution models \cite{ClaretTorres16aa,Tkachenko+20aa}.  

The volume and quality of light curve data has increased enormously in recent years \cite{Me21univ}, especially from space-based exoplanet surveys such as \corot\ 
\cite{Deleuil+18aa} and NASA's \emph{Kepler} \cite{Borucki16rpp}, K2 \cite{Howell+14pasp} and TESS \cite{Ricker+15jatis} (Transiting Exoplanet Survey Satellite) missions. This work is one of a series where we revisit known dEBs in order to refine their characterisation with the benefit of this new era of photometry. Here we analyse \targfull\ using TESS light curves alongside previously published RV data.


\section*{The dEB \targfull}

HD 19115 was categorised as photometrically variable in 1967 by Strohmeier and Ott  \cite{Strohmeier+67ibvs}. Popper \cite{Popper83aj} reported that the spectrum was double-lined and it was given the designation \targ\ by Kukarkin et al.\ \cite{Kukarkin+72ibvs}. Chen \cite{Chen75acta} reported on \emph{UBV} photometric observations made at the Rosemary Hill Observatory between 1970 and 1972, publishing an ephemeris and relative sizes for the components. Further photometric observations were made by Mauder and Ammann \cite{MauderAmmann76} and, with the addition of spectroscopic observations made available to them by Popper, they recorded masses and radii for both components to 2--3\% confidence and assigned a spectral type of F0.  

Popper and Dumont \cite{PopperDumont77aj} included \targ\ in their program of \emph{UBV} photometric observations at the Palomar and Kitt Peak observatories with the \emph{B}- and \emph{V}-band magnitudes given in Table \ref{tab:info} being recorded over 11 nights.  Brancewicz \& Dworak \cite{BrancewiczDworak80acta} included it in their catalogue of eclipsing binaries where they used numerical methods to characterise the system, giving a spectral type of F0+ and physical parameters to $\sim$5\% confidence.

The most complete investigations were carried out by Popper \cite{Popper83aj,Popper80araa} based on spectrograms taken at the Lick Observatory between 1967 and 1974 along with photometry from Chen \cite{Chen75acta} and Mauder \& Ammann \cite{MauderAmmann76}. Popper determined the spectral types of the components as F1 and F4, gave their masses to 2\% confidence, and determined their radii to 2.5\% and 4.5\% for the primary and secondary components respectively.

Outside automated surveys, in the years since Popper few observations have been made. Wolf \& Kern \cite{WolfKern83apjs} recorded three observations as part of their photometric survey of the southern hemisphere, giving a \emph{V}-band magnitude ranging from $8.39$ at quadrature to $8.90$ during primary eclipse. Perry \& Christodoulou \cite{PerryChristodoulou96pasp} included it in their \emph{uvby$\beta$} interstellar reddening survey of the southern hemisphere. Nordstr\"om et al.\ \cite{Nordstrom+97aas} made three spectroscopic observations as part of their RV survey of early F-type dwarfs. 

\begin{table}[t]
\caption{\em Basic information on \targ. \label{tab:info}}
\centering
\begin{tabular}{lll}
{\em Property}                            & {\em Value}                 & {\em Reference}                   \\[3pt]
Right ascension (J2000)                   & 03:03:59.95                 & \cite{Gaia21aa}                   \\
Declination (J2000)                       & $-$17:44:16.06              & \cite{Gaia21aa}                   \\
Henry Draper designation                  & HD 19915                    & \cite{CannonPickering20anhar}     \\
\textit{Hipparcos} designation            & HIP 14273                   & \cite{Hipparcos97}                \\
\textit{Tycho} designation                & TYC 5868-881-1              & \cite{Hog+00aa}                   \\
\textit{Gaia} DR3 designation             & 5152756553745197952         & \cite{Gaia21aa}                   \\
\textit{Gaia} DR3 parallax                & $5.2380 \pm 0.0198$ mas     & \cite{Gaia21aa}                   \\  
TESS\ Input Catalog designation           & TIC 98853987                & \cite{Stassun+19aj}               \\
$B$ magnitude                             & $8.79 \pm 0.07$             & \cite{PopperDumont77aj}           \\  
$V$ magnitude                             & $8.43 \pm 0.07$             & \cite{PopperDumont77aj}           \\
$G$ magnitude                             & $8.306 \pm 0.003$           & \cite{Gaia21aa}                   \\
$J$ magnitude                             & $7.799 \pm 0.020$           & \cite{Cutri+03book}               \\	
$H$ magnitude                             & $7.659 \pm 0.034$           & \cite{Cutri+03book}               \\
$K_s$ magnitude                           & $7.626 \pm 0.023$           & \cite{Cutri+03book}               \\
Spectral type                             & F2 V                        & \cite{HoukSmith-Moore88}          \\[3pt]
\end{tabular}
\end{table}


Table~\ref{tab:info} shows basic information for \targ.  The \emph{B} and \emph{V} magnitudes are those recorded by Popper and Dumont \cite{PopperDumont77aj}. These were explicitly based on observations made outside of an eclipse and have since been widely used. The \emph{J}, \emph{H} and \emph{$K_s$} magnitudes are those reported by 2MASS from observations made at JD $2\,451\,052.9027\pm30\sec$. At this time the system will have been within a secondary eclipse so these will be below the system's maximum brightness. The spectral type of F2~V is given by Houk \& Smith-Moore \cite{HoukSmith-Moore88} as part of the Michigan Catalogue of HD Stars, Vol 4.


\section*{Observational material}

\begin{figure}[t] \centering 
	\includegraphics[width=\textwidth]{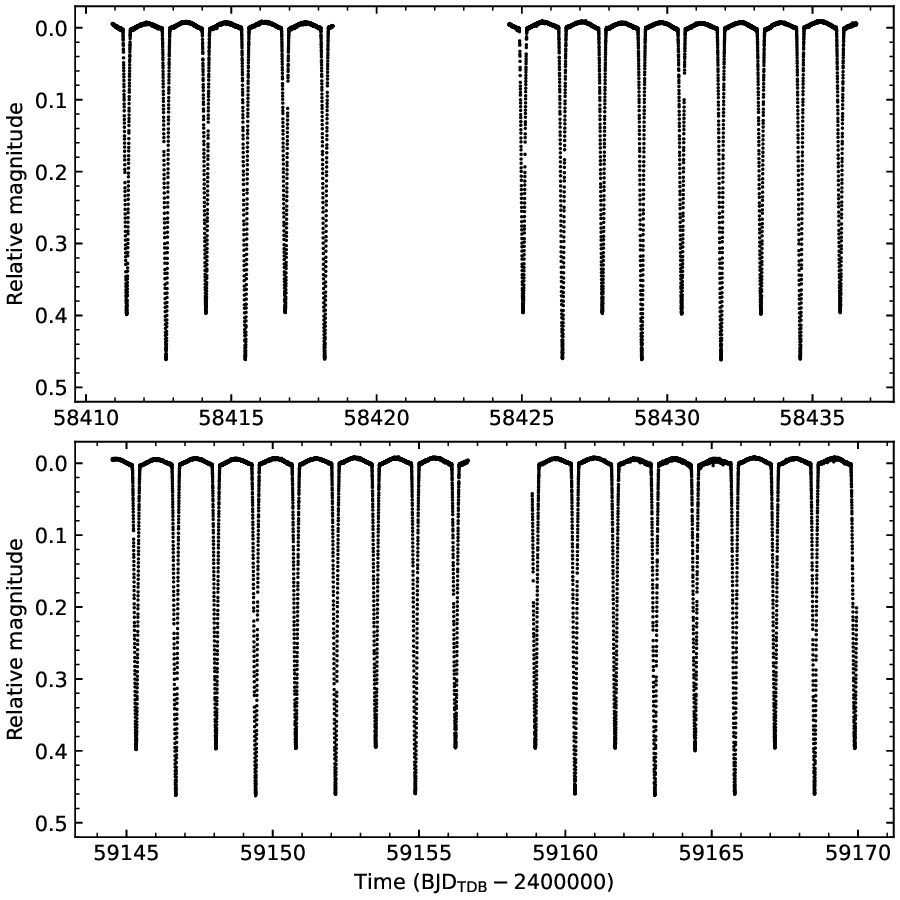} \\
\caption{\label{fig:time} TESS short-cadence SAP photometry of \targ\ from sectors 4
(top) and 31 (bottom). The flux measurements have been converted to magnitude units then
rectified to zero magnitude by the subtraction of low-order polynomials.} \end{figure}


\targ~was observed twice by the TESS mission \cite{Ricker+15jatis}, first in sector 4 from 2018/10/19 to 2018/11/14 and again in sector 31 from 2020/10/22 to 2020/11/18, each in short cadence mode with a $120$~s sampling rate. Both sectors show light curves covering a period of approximately 25 days with a break near the midpoint for data download. Unambiguous primary and secondary eclipses are seen in addition to a sinusoidal variation resulting from the ellipsoidal effect (Fig.~\ref{fig:time}).

The TESS time series data for the two sectors were downloaded from the MAST archive\footnote{Mikulski Archive for Space Telescopes, \\ \texttt{https://mast.stsci.edu/portal/Mashup/Clients/Mast/Portal.html}} and subsequently processed using the {\sc lightkurve} \cite{Lightkurve+18} and {\sc astropy} \cite{Astropy22apj} Python packages. These data consist of simple aperture photometry (SAP) and pre-search data conditioning SAP (PDCSAP) flux measurements \cite{Jenkins+16spie}.  We based our analysis on the SAP data as it is well-behaved whereas extraneous variability was seen in the PDCSAP data from sector 4. Data points with no flux value recorded (NaN) and those with a non-zero QUALITY flag were cut, as were those within a distorted secondary eclipse within sector 4 from BJD $2\,458\,420.0$ to $2\,458\,423.0$. A total of $13\,841$ data points from sector 4 and $16\,671$ from sector 31 were considered for subsequent analysis.

The \gaia\ DR3 database\footnote{\texttt{https://vizier.cds.unistra.fr/viz-bin/VizieR-3?-source=I/355/gaiadr3}} was queried for potential sources of third light within 2~arcmin of \targ. Six of the seven objects found are at least 10~mag fainter than \targ\ in the \emph{G}-band so contribute negligible light. The remaining object, TYC 5868-428-1, has a \emph{G}-band magnitude of 11.053~mag with the resulting flux ratio of 0.080 being adopted as the initial value of the fitted third light parameter in the following analysis.



\begin{figure}[t] \centering
	\includegraphics[width=\textwidth]{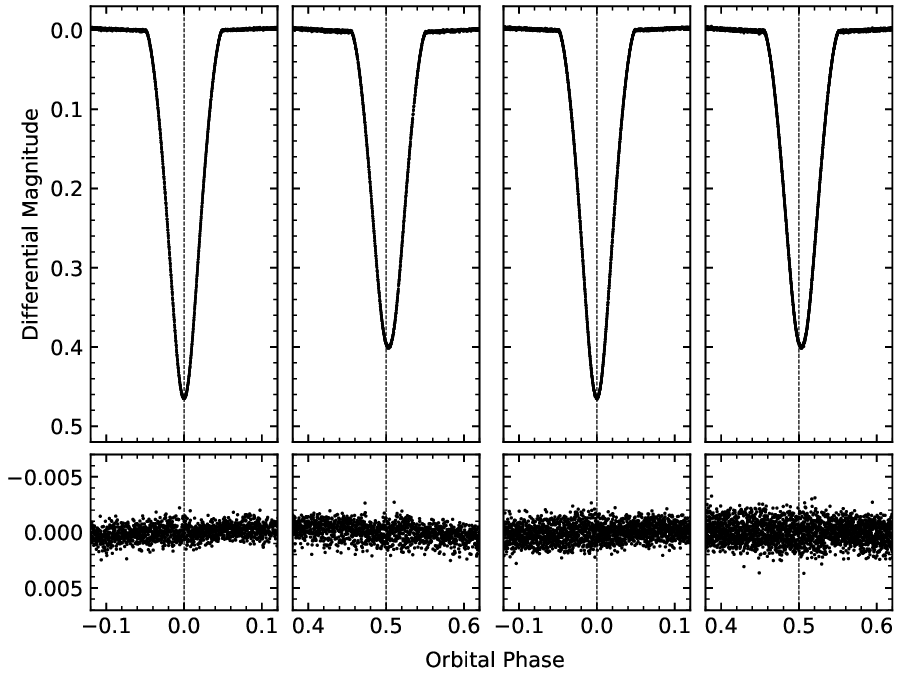} \\
	\caption{\label{fig:phase-s0004} Best fit to the full TESS\ sector 4 light curve of \targ\ using
		{\sc jktebop}. The primary and secondary eclipse of the first half-sector are shown to the left 
		and those for the second half-sector to the right. The residuals are shown on an enlarged 
		scale in the lower panels.}
\end{figure}

\section*{Light curve analysis}

The remaining SAP flux data were converted to magnitudes then rectified to zero and detrended by fitting and subtracting a quadratic polynomial across the whole of each sector. This was refined after initial attempts at fitting, with the best results achieved by subtracting a second quadratic fit from those data in sector 4 following the mid-sector break. The resulting light curves, shown in Fig.~\ref{fig:time}, consist of four isolated half-sectors over a time interval of $\sim$759~d. We adopt the standard definition of the primary eclipse as being the deeper of the two which occurs when the larger and brighter component, which we label star A, is eclipsed by the smaller star B.

\begin{figure}[t] \centering
	\includegraphics[width=\textwidth]{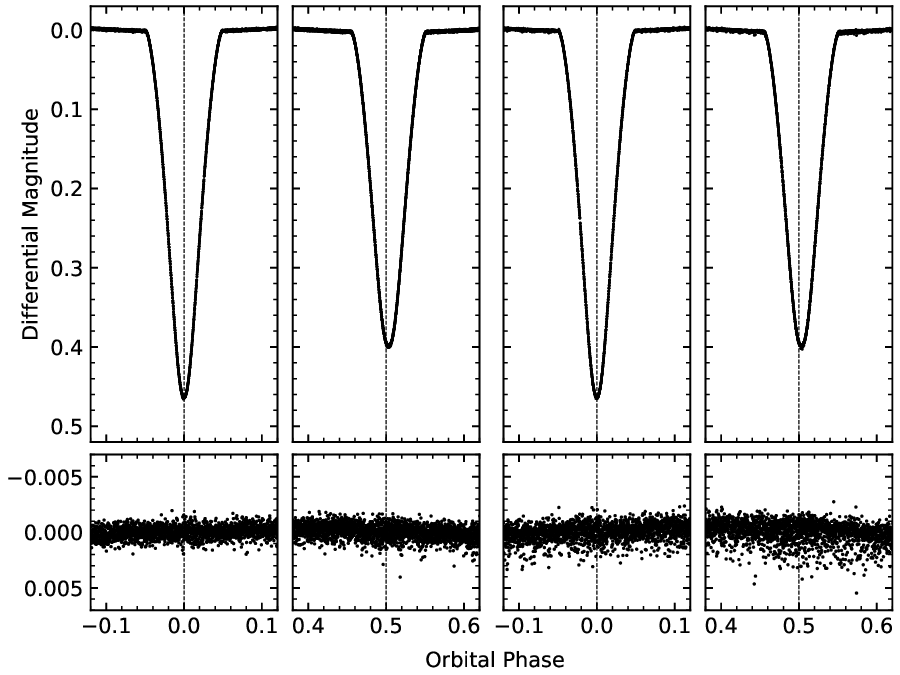} \\
	\caption{\label{fig:phase-s0031} Same as Fig. \ref{fig:phase-s0004} 
		but for TESS data from sector 31.}
\end{figure}

The data were fitted using version 43 of the {\sc jktebop}\footnote{\texttt{http://www.astro.keele.ac.uk/jkt/codes/jktebop.html}} code \cite{Me++04mn2,Me13aa} with a total $30\,512$ datapoints fitted as the four separate half-sectors. Each light curve was fitted for the orbital period ($P$) and the time of mid-primary eclipse ($T_0$) with our reference time being the primary eclipse closest to the midpoint of the data.  Also fitted were the sum ($r_{\rm A}+r_{\rm B}$) and ratio ($k=r_{\rm B}/r_{\rm A}$) of the fractional radii, the orbital inclination ($i$), the orbital eccentricity ($e$) and argument of periastron ($\omega$) through the Poincar\'e elements $e \cos{\omega}$ and $e \sin{\omega}$, the stars' central surface brightness ratio ($J$), the amount of third light ($L_3$) and each star's reflected light.  

\begin{sidewaystable} \centering
	\caption{\em \label{tab:jktebop-ext} The fitted parameters of \targ\ for each of the four TESS
		half-sector light curves using the {\sc jktebop} code. The uncertainties are 1$\sigma$ values
		derived from either Monte Carlo or residual-permutation simulations. For each parameter the
		uncertainties given are from the method yielding the larger weighted mean uncertainty
		across the half-sectors. 2400000 has been subtracted from the times to save space.
}
\begin{tabular}{lcccc}
	{\em Parameter}                           &       {\em Sector 4.1}            &       {\em Sector 4.2}            &       {\em Sector 31.1}           &       {\em Sector 31.2}           \\[3pt]
	{\it Fitted parameters:} \\
	Primary eclipse time (BJD$_{\rm TDB}$)    & $   58415.482929 \pm   0.000022 $ & $   58431.853144 \pm   0.000015 $ & $   59152.142894 \pm   0.000015 $ & $   59163.056373 \pm   0.000020 $ \\     
	Orbital period (d)                        & $      2.7283891 \pm  0.0000157 $ & $      2.7283696 \pm  0.0000092 $ & $      2.7283737 \pm  0.0000065 $ & $      2.7283805 \pm  0.0000117 $ \\     
	Orbital inclination (\degr)               & $         86.366 \pm      0.037 $ & $         86.373 \pm      0.033 $ & $         86.412 \pm      0.022 $ & $         86.313 \pm      0.040 $ \\     
	Sum of the fractional radii               & $        0.30659 \pm    0.00021 $ & $        0.30676 \pm    0.00019 $ & $        0.30651 \pm    0.00013 $ & $        0.30682 \pm    0.00023 $ \\     
	Ratio of the radii                        & $         0.7042 \pm     0.0017 $ & $         0.7033 \pm     0.0011 $ & $         0.7048 \pm     0.0010 $ & $         0.7026 \pm     0.0012 $ \\     
	Central surface brightness ratio          & $         0.9262 \pm     0.0065 $ & $         0.9203 \pm     0.0061 $ & $         0.9309 \pm     0.0041 $ & $         0.9203 \pm     0.0068 $ \\     
	Third light                               & $        -0.0025 \pm     0.0019 $ & $        -0.0014 \pm     0.0017 $ & $         0.0024 \pm     0.0012 $ & $        -0.0036 \pm     0.0021 $ \\     
	LD $c$ coefficient of star~A              & $          0.592 \pm      0.027 $ & $          0.622 \pm      0.025 $ & $          0.573 \pm      0.017 $ & $          0.611 \pm      0.028 $ \\     
	LD $c$ coefficient of star~B              & $          0.614 \pm      0.019 $ & $          0.601 \pm      0.018 $ & $          0.620 \pm      0.012 $ & $          0.608 \pm      0.021 $ \\     
	LD $\alpha$ coefficient of star~A         & \multicolumn{4}{c}{$0.4676$ (fixed)} \\     
	LD $\alpha$ coefficient of star~B         & \multicolumn{4}{c}{$0.4967$ (fixed)} \\     
	$e \cos{\omega}$                          & $        0.00492 \pm    0.00002 $ & $        0.00491 \pm    0.00001 $ & $        0.00513 \pm    0.00001 $ & $        0.00513 \pm    0.00002 $ \\     
	$e \sin{\omega}$                          & $       -0.01181 \pm    0.00108 $ & $       -0.01290 \pm    0.00069 $ & $       -0.01135 \pm    0.00066 $ & $       -0.01229 \pm    0.00082 $ \\     
	{\it Derived parameters:} \\
	Fractional radius of star~A               & $        0.17990 \pm    0.00026 $ & $        0.18010 \pm    0.00017 $ & $        0.17979 \pm    0.00016 $ & $        0.18021 \pm    0.00018 $ \\     
	Fractional radius of star~B               & $        0.12669 \pm    0.00016 $ & $        0.12666 \pm    0.00011 $ & $        0.12672 \pm    0.00010 $ & $        0.12661 \pm    0.00012 $ \\     
	Orbital eccentricity                      & $        0.01279 \pm    0.00099 $ & $        0.01381 \pm    0.00064 $ & $        0.01245 \pm    0.00060 $ & $        0.01332 \pm    0.00075 $ \\     
	Light ratio $\ell_{\rm B}/\ell_{\rm A}$   & $         0.4535 \pm     0.0014 $ & $         0.4537 \pm     0.0010 $ & $         0.4542 \pm     0.0009 $ & $         0.4510 \pm     0.0010 $ \\[3pt]
\end{tabular}
\end{sidewaystable}

We adopted the power-2 limb darkening (LD) law with TESS-specific coefficients taken from Claret \& Southworth \cite{ClaretSouthworth22aa}. The coefficients were interpolated for star~A ($\Teff = 6840\,\mathrm{K}$ and $\log{g}=4.0$) and star~B ($\Teff = 6560\,\mathrm{K}$ and $\log{g}=4.2$), each with a solar metallicity ($\mathrm{[Fe/H]}=0.0$). For both stars, the scaling coefficient $c$ was left free to fit and $\alpha$ was fixed. 

The best fits to the light curves are shown in Figs.~\ref{fig:phase-s0004} and \ref{fig:phase-s0031} where it can be seen that the secondary eclipse is slightly offset from phase 0.50, confirming a small orbital eccentricity. As F-type stars may exhibit $\gamma$\,Doradus or $\delta$\,Scuti pulsations \cite{UytterhoevenMoya+11a&a} the residuals of the fits were analysed with Lomb-Scargle periodograms, but no evidence of pulsation was found.

\begin{table} \centering
	\caption{\em \label{tab:jktebop} The adopted parameters of \targ\ derived from the four TESS half-sector light curves fitted with the {\sc jktebop} code. Other than the time of primary eclipse each is the weighted mean of the corresponding fitted parameter values and 1$\sigma$ uncertainties for each fitted half-sector given in Table.~\ref{tab:jktebop-ext}.
	}
	\begin{tabular}{lc}
		{\em Parameter}                           &       {\em Value}                 \\[3pt]
		{\it Fitted parameters:} \\
		Time of primary eclipse (BJD$_{\rm TDB}$) & $ 2458415.482929 \pm   0.000022 $ \\     
		Orbital period (d)                        & $      2.7283751 \pm  0.0000068 $ \\     
		Orbital inclination (\degr)               & $         86.381 \pm      0.042 $ \\     
		Sum of the fractional radii               & $        0.30662 \pm    0.00015 $ \\     
		Ratio of the radii                        & $         0.7037 \pm     0.0011 $ \\     
		Central surface brightness ratio          & $         0.9262 \pm     0.0057 $ \\     
		Third light                               & $        -0.0002 \pm     0.0030 $ \\     
		LD $c$ coefficient of star~A              & $          0.593 \pm      0.024 $ \\     
		LD $c$ coefficient of star~B              & $          0.613 \pm      0.009 $ \\     
		LD $\alpha$ coefficient of star~A         & $         0.4676$ (fixed)         \\     
		LD $\alpha$ coefficient of star~B         & $         0.4967$ (fixed)         \\     
		$e \cos{\omega}$                          & $        0.00502 \pm    0.00013 $ \\     
		$e \sin{\omega}$                          & $       -0.01210 \pm    0.00076 $ \\     
		{\it Derived parameters:} \\
		Fractional radius of star-A               & $        0.18000 \pm    0.00020 $ \\     
		Fractional radius of star-B               & $        0.12667 \pm    0.00005 $ \\     
		Orbital eccentricity                      & $        0.01310 \pm    0.00067 $ \\     
		Light ratio $\ell_{\rm B}/\ell_{\rm A}$   & $         0.4532 \pm     0.0015 $ \\[3pt]
	\end{tabular}
\end{table}

The final values and uncertainties for the fitted parameters of each half-sector were separately determined with 10,000 Monte Carlo (MC) simulations \cite{Me++04mn2} and a residual permutation (RP) algorithm \cite{Me08mn}, as implemented by {\sc jktebop} tasks 8 and 9 respectively. The latter method successively shifts the best-fit residuals along the light curve until they are cycled back to their initial position. With each shift a new fit is made and the final distribution of each fitted parameter gives an estimate of its uncertainty. While the MC simulations are sensitive to Poisson noise, the RP algorithm is additionally sensitive to correlated noise \cite{Me08mn}. The fitted parameters for each half-sector are given in Table~\ref{tab:jktebop-ext} with the uncertainties being the 1$\sigma$ values of either the MC or RP simulations. The selection of uncertainties for each parameter is based on the method which yields the larger weighted mean errorbar. The adopted parameters for \targ, as given in Table~\ref{tab:jktebop}, are the weighted mean and uncertainty of the corresponding fitted parameters across the four half-sectors.


\section*{Orbital ephemeris}

\begin{table} \centering
\caption{\em Times of published mid-eclipse for \targ\ and their residuals versus the fitted ephemeris.\label{tab:tmin}}
\setlength{\tabcolsep}{10pt}
\begin{tabular}{rr@{.}lr@{.}lr@{.}ll}
{\em Orbital} & \mc{\em Eclipse time} & \mc{\em Uncertainty} & \mc{\em Residual} & {\em Reference} \\
{\em cycle}   & \mc{\em (BJD$_{TDB}$)}    & \mc{\em (d)}          & \mc{\em (d)}     &              \\[3pt]
$-4130.5$ & 2441230&842503 &   ~~~0&000400 &   $-$0&000439 & \cite{Chen75acta}       \\   
$-4121.0$ & 2441256&762604 &      0&000400 &   $ $0&000144 & \cite{Chen75acta}       \\   
$-4117.0$ & 2441267&676804 &      0&000400 &   $ $0&000863 & \cite{Chen75acta}       \\   
$-4110.0$ & 2441286&774304 &      0&000400 &   $-$0&000228 & \cite{Chen75acta}       \\   
$-4106.0$ & 2441297&687604 &      0&000400 &   $-$0&000409 & \cite{Chen75acta}       \\   
$-4100.5$ & 2441312&692704 &      0&000400 &   $-$0&001345 & \cite{Chen75acta}       \\   
$-3986.5$ & 2441623&728610 &      0&000400 &   $ $0&000352 & \cite{Chen75acta}       \\   
$-3981.0$ & 2441638&734809 &      0&000400 &   $ $0&000515 & \cite{Chen75acta}       \\   
$-3967.5$ & 2441675&567908 &      0&000400 &   $ $0&000616 & \cite{Chen75acta}       \\   
$-3966.0$ & 2441679&659808 &      0&000400 &   $-$0&000039 & \cite{Chen75acta}       \\   
$ 2168.0$ & 2458415&482929 &      0&000109 &   $ $0&000003 & This work               \\   
$ 2174.0$ & 2458431&853144 &      0&000075 &   $-$0&000004 & This work               \\   
$ 2438.0$ & 2459152&142894 &      0&000074 &   $ $0&000002 & This work               \\   
$ 2442.0$ & 2459163&056373 &      0&000099 &   $-$0&000000 & This work               \\   
\end{tabular}
\end{table}


With the light curve analysis yielding consistent orbital parameters we sought to derive a high-precision orbital ephemeris for the system.  In order to base this on the longest possible dataset, historical minima timing data for \targ\ were obtained from the TIming DAtabase at Krakow (TIDAK) team \cite{Kreiner04acta}. While the majority of minima timings were given without an uncertainty, all included a weight value between 1 and 10. Where missing, estimated uncertainties were generated by scaling a base estimate of $0.004$ by the reciprocal of the observation's weight.  To these data were added the primary epoch and period from fitting each of the four half-sectors of TESS data with their uncertainties scaled up by a factor of 5 to cover any scatter.

\begin{figure}[t] \centering \includegraphics[width=\textwidth]{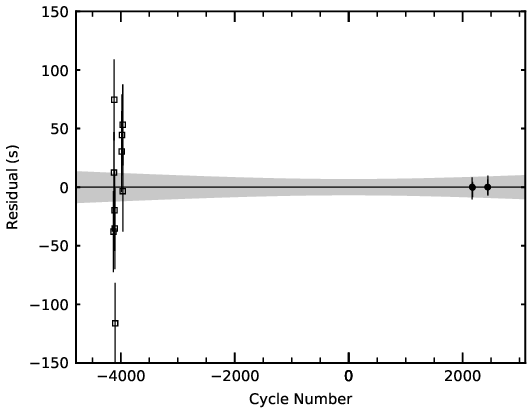} \\
\caption{\label{fig:ephemresidual} Observed minus calculated ($O-C$) diagram of the times of primary
	minimum versus the fitted linear ephemeris. Timings from the TESS data are shown with with filled circles
	and those from the literature are shown as open squares where uncertainties have been estimated. 
	The shaded areas indicate the $1\sigma$ uncertainty in the ephemeris determined from these data.} 
\end{figure}

The existing TIDAK ephemeris\footnote{\texttt{https://www.as.up.krakow.pl/minicalc/ERICW.HTM}} was used to calculate cycle numbers and assign minima types (primary or secondary) to the eclipses after which linear, quadratic and cubic polynomials were fitted to reveal trends in the timings. Initial attempts at fitting the data revealed excessive scatter from a number of sources and the final fitted ephemeris is based only on the TESS observations and those from TIDAK with a weight of 10. With the fitting complete the quadratic and cubic fits were discounted, as they were poorly constrained by the data, and the following linear ephemeris was adopted: 
\begin{equation}
  \mbox{Min~I} = {\rm BJD}_{\rm TDB}~ 2452500.37624 (69) + 2.72837024 (27) E
\end{equation}
with $E$ being the cycle number since the reference time and the bracketed values being the uncertainties in the last digit of the preceding values. The final eclipse timing data used in this analysis are given in Table \ref{tab:tmin} and the residuals of the linear fit are shown in Fig.~\ref{fig:ephemresidual}.

\section*{Radial velocities}

\begin{figure}[t] \centering \includegraphics[width=\textwidth]{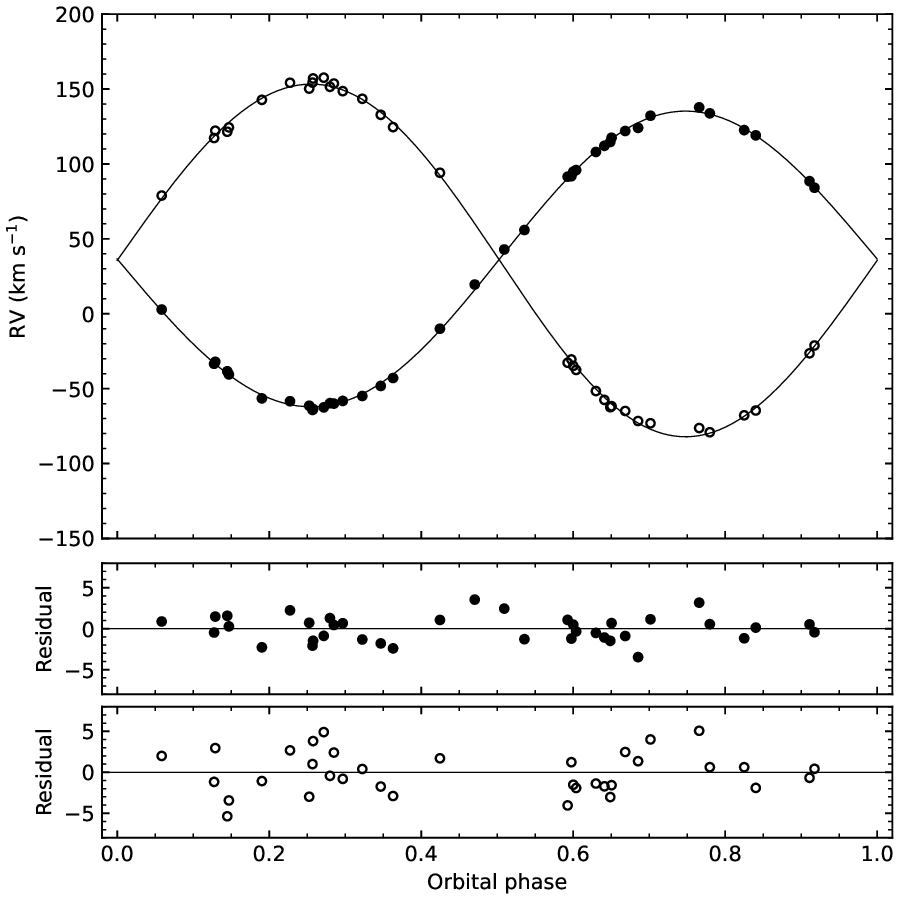} \\
\caption{\label{fig:popper} RVs of \targ\ measured by Popper \cite{Popper83aj}
(filled circles for star~A and open circles for star~B) compared to the best-fitting
spectroscopic orbits from {\sc jktebop} (solid curves). The residuals are given in
the lower panels separately for the two components.} 
\end{figure}

The RV measurements originally published by Popper \cite{Popper83aj} were reanalysed.  The observations were made between 1967 and 1974 at the Lick Observatory and consist of 38 RVs for star A and 35 for star B. Popper's data gives RVs to one decimal place and HJD timestamps to three decimal places and, in the absence of uncertainties, we applied equal weighting to all measurements.  The RVs were analysed with {\sc jktebop} based on the ephemeris and orbital parameters derived from the photometric fitting with the uncertainties of the fitted results determined using Monte Carlo simulations (see Paper\,VI, ref. \cite{Me21obs5}).

Initial fitting was carried out with fixed values for $T_0$ and $P$ which yielded results very similar to Popper's with slightly worse rms residuals for star B. Given the low temporal resolution of the observations, we investigated whether allowing these parameters to be varied when fitting the RV orbits would yield an improved fit.  It was found that allowing $T_0$ to vary yielded a demonstrable improvement in the fitted RV orbits with lower uncertainties and rms residuals; these are the results reported here.

The fitted orbits are shown in Fig.~\ref{fig:popper}.  Parameters for the spectroscopic orbits are given in Table~\ref{tab:orbits} which shows them to be in good agreement with Popper \cite{Popper83aj} while having lower uncertainties and residuals. Nordstr\"om et al.\ \cite{Nordstrom+97aas} give only an overall systemic velocity of $\Vsys=37.16\pm2.94\,\mathrm{km\,s^{-1}}$, based on three observations, which also agrees well with our findings. Few works have published any further RV data on \targ, with Duflot et al.\ \cite{Duflot+95aas} giving $\Vsys=36.4\,\mathrm{km\,s^{-1}}$ in their Wilson-Evans-Batten catalogue and Gontcherov \cite{Gontcharov06al} giving a value $\Vsys=36.8\pm2.1\,\mathrm{km\,s^{-1}}$, potentially based on the values published by Nordstr\"om and Duflot, with both showing some overlap with our individual RVs.

\begin{table} \centering
\caption{\em \label{tab:orbits} Spectroscopic orbits for \targ\ from the literature
and from the reanalysis of the RVs in the current work. All quantities are in\kms.}
\begin{tabular}{lrrrrc}
	{\em Source}        & $K_{\rm A}$~ & $K_{\rm B}$~ & ${\Vsys}_{\rm ,A}$~ & ${\Vsys}_{\rm ,B}$~ & $rms$ residual \\[7pt]
	Popper \cite{Popper83aj}                     &       98.9 &      118.0 &       36.4 &       35.7 & 1.70, 2.80   \\
												& $\pm$  0.3 & $\pm$  0.6 & $\pm$  0.3 & $\pm$  0.5 &              \\     
	This work                                   &       98.7 &      117.7 &       36.1 &       36.2 & 1.55, 2.55   \\
												& $\pm$  0.3 & $\pm$  0.5 & $\pm$  0.3 & $\pm$  0.4 &              \\[7pt]
\end{tabular}
\end{table}

%


\section*{Physical properties of \targ}

The physical properties of \targ\ were calculated based on the parameters derived from the light curves (Table~\ref{tab:jktebop}), the RV fitting (Table~\ref{tab:orbits}) and the new ephemeris calculated above. The uncertainties for $r_{\rm A}$ and $r_{\rm B}$ were increased to 0.2\% following the  recommendation from Maxted et al.\ \cite{Maxted+20mnras}. Effective temperature values were taken from Popper \cite{Popper83aj} where a value for both components has been given with accompanying uncertainty. The {\sc jktabsdim} code \cite{Me++05aa} was used to calculate the system's properties given in Table~\ref{tab:absdim} with uncertainties propagated using a perturbation approach. Standard formulae \cite{Hilditch01book} and the reference solar values from the IAU \cite{Prsa+16aj} were used. 

The results show that the masses and radii are determined to a precision of better than 1.0\%, meeting the criteria for inclusion in the Detached Eclipsing Binary Catalogue (DEBCat\footnote{\texttt{https://www.astro.keele.ac.uk/jkt/debcat/}} Ref. \cite{Me15aspc}). The mass measurements are in agreement with the original values published by Popper \cite{Popper83aj}, as expected as they are based on the same RV data. The measured radii are consistent with those from Popper ($2.08\pm0.05$ and $1.56\pm0.07$\Rsun) but are much more precise due to the availability of the TESS photometry. 

We determined the distance to the system based on the $B$ and $V$ apparent magnitudes from Popper \& Dumont \cite{PopperDumont77aj} and those in the $J$, $H$ and $K_s$-bands from 2MASS \cite{Cutri+03book} (Table~\ref{tab:info}). The 2MASS magnitudes are based on observations made during a secondary eclipse which were corrected for by subtracting the fitted light curve model at the corresponding phase to find revised values of $J=7.658\pm0.022$, $H=7.518\pm0.035$ and $K_s=7.485\pm0.025$~mag. We searched for reliable observations made in the Cousins $R$ and $I$-band but found none. An interstellar extinction value of $\EBV=0.013\pm0.015$ was adopted from the {\sc stilism} tool \footnote{\texttt{https://stilism.obspm.fr}} and bolometric corrections from Girardi et al.\ \cite{Girardi+2002aa} were used. 

\begin{table} \centering
	\caption{\em Physical properties of \targ\ defined using the nominal solar units given by IAU
		2015 Resolution B3 (ref.\ \cite{Prsa+16aj}). \label{tab:absdim}}
\begin{tabular}{lr@{\,$\pm$\,}lr@{\,$\pm$\,}l}
	{\em Parameter}               & \multicolumn{2}{c}{\em Star A} & \multicolumn{2}{c}{\em Star B}    \\[3pt]
	Mass ratio                                   &   \multicolumn{4}{c}{$   0.8385 \pm   0.0047 $}     \\       
	Semimajor axis of relative orbit (\Rsunnom)  &   \multicolumn{4}{c}{$   11.694 \pm    0.034 $}     \\       
	Mass (\Msunnom)                              &    1.568 &    0.016      &    1.314 &    0.010      \\       
	Radius (\Rsunnom)                            &   2.1048 &   0.0074      &   1.4812 &   0.0052      \\       
	Surface gravity ($\log$[cgs])                &   3.9869 &   0.0026      &   4.2156 &   0.0022      \\       
	Density ($\!\!$\rhosun)                      &   0.1681 &   0.0011      &   0.4045 &   0.0027      \\       
	Effective temperature (K)                    &     6839 &       87      &     6561 &       98      \\       
	Luminosity $\log(L/\Lsunnom)$                &    0.941 &    0.022      &    0.564 &    0.026      \\       
	$M_{\rm bol}$ (mag)                          &    2.387 &    0.056      &    3.330 &    0.065      \\       
	Distance (pc)                                &   \multicolumn{4}{c}{$    191.7 \pm      3.8 $}     \\       
\end{tabular}
\end{table}


The \gaia\ DR3 \cite{Gaia21aa} parallax yields a distance of $187.9^{+0.6}_{-0.9}$~pc for \targ, with the renormalized unit weight error (RUWE) of 1.029 indicating a reliable astrometric solution. This is in agreement with distances based on the bolometric corrections of Girardi et al.\ \cite{Girardi+2002aa} when applied to the $B$, $V$ and $J$-band magnitudes, with the $J$-band yielding the best match at $188.4\pm3.0$~pc. A similar pattern is seen when using the passband-specific surface brightness-\Teff\ relations of Kervella et al.\ \cite{Kervella+04aa}. Both methods yield slightly larger distances than \gaia\ using the $H$ and $K_s$-band magnitudes, however the weighted mean of all of the derived distances is $191.7\pm3.8$~pc which is in agreement with the \gaia\ value.



\section*{Comparison with theoretical models}

\begin{figure}[t] \centering \includegraphics[width=\textwidth]{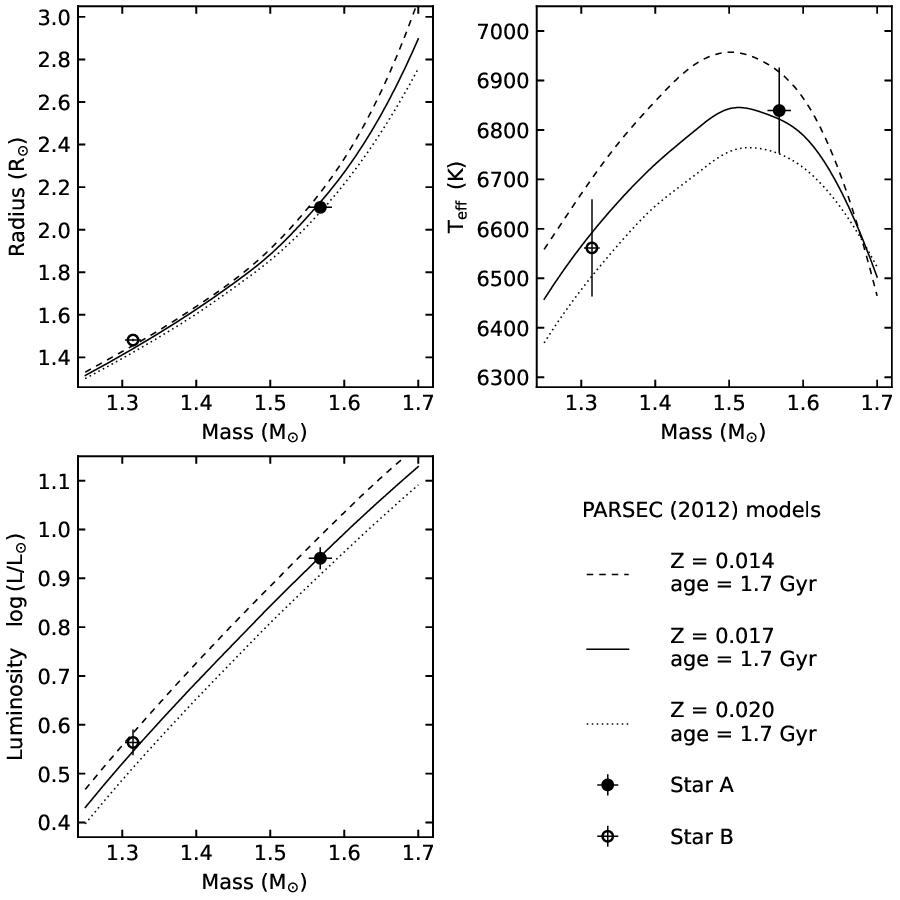} \\
\caption{\label{fig:models} Comparison between the theoretical predictions of \textsc{parsec} models \cite{Bressan+12mnras}
	and the measured properties of \targ\ presented here for stellar mass versus radius, $\mathrm{T_{eff}}$ and luminosity.~
	The ages and metal abundances of the chosen theoretical models are given in the legend within the lower right quadrant.} 
\end{figure}

To test our results, the measured properties of the components of \targ\ were compared with predictions of \textsc{parsec} theoretical stellar evolutionary models \cite{Bressan+12mnras} in plots of mass versus radius, \Teff\ and luminosity. The best agreement was found for models based on a solar composition (fractional metal abundance of $Z=0.017$) and an age of 1.7~Gyr. This gives a very good fit to star A with star B appearing slightly larger and more luminous than the model predictions. Choosing a model with lower metallicity gives a closer fit to star B's radius and luminosity but at the expense of star A which now appears slightly smaller and less massive than predicted and both components are cooler than the model. The converse is found when higher-metallicity models are used with the \Teff\ being most sensitive to any change. Fig.~\ref{fig:models} shows models ranging from slightly sub-solar ($Z=0.014$) through solar to slightly super-solar metal abundance ($Z=0.020$). 

\begin{figure}[t] \centering \includegraphics[width=\textwidth]{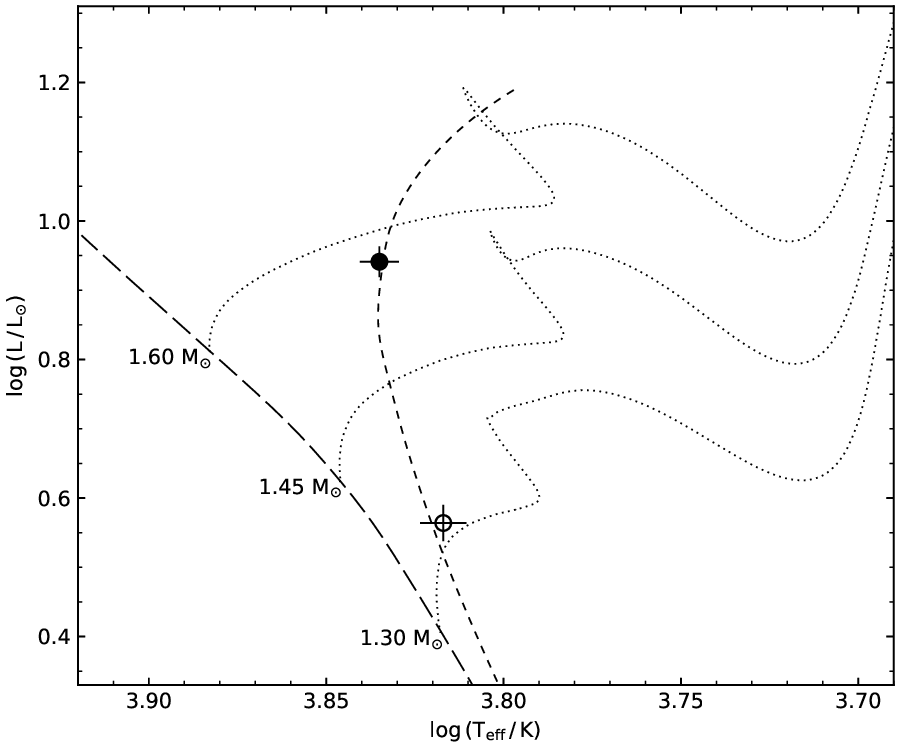} \\
\caption{\label{fig:hrd} Hertzsprung-Russell diagram showing the components of \targ\
	(filled circle for star A and open circle for star B) and selected predictions from the \textsc{parsec} models \cite{Bressan+12mnras}
	(dotted lines). The zero-age main sequence is shown with a long-dashed line, and an ioschrone for an age of 1.7~Gyr with a short-dashed line. 
	Models for 1.30, 1.45 	and 1.60\Msun\ are shown (labelled) with a metal abundance of $Z=0.017$.} 
\end{figure}

This choice of age and metallicity is further supported with a Hertzsprung-Russell diagram showing the evolutionary tracks of \textsc{parsec} model stars of $Z=0.017$ and masses 1.3, 1.45 and 1.6\Msun\ (Fig.~\ref{fig:hrd}).  Both components of \targ\ are consistent with model stars of similar mass which have evolved away from the ZAMS line into the upper half of the main sequence.  Comparisons were also made with equivalent MIST \cite{Dotter16apjs:MIST,Choi+16apj:MIST} models and evolutionary tracks, with broadly similar results except that the star A is hotter than predicted.

%

While the chosen \textsc{parsec} model has a good fit to the mass, radii and \Teff\ of the components, we note that the metallicity is in disagreement with published values. Perry \& Christodoulou \cite{PerryChristodoulou96pasp} give $[\mathrm{Fe/H}] = -0.32$ in their $uvby\beta$ survey of southern hemisphere A and F-type stars. A value of $[\mathrm{Fe/H}]=-0.39$ was published in the Geneva-Copenhagen Catalogue by Holmberg et al.\ \cite{Holmberg+07aa} and subsequently recalculated as $-0.26$ by Casagrande et al.\ \cite{CasagrandeSchonrich+11aa}. A plausible answer to this discrepancy is that the metallicity was calculated assuming the photometry of the system represents that of a single star rather than the combined light of two stars of different colours.

\section*{Summary}

\targ\ is a dEB consisting of a pair of F-type stars that has remained largely ignored since it was last studied by Popper in 1983 \cite{Popper83aj}.  We have revisited the system, making use of two sectors of TESS photometry, and determined its photometric parameters to high precision. By combining these results with Popper's original RVs we refined the spectroscopic orbits and subsequently obtained high-quality measurements of the physical properties of the system. The residuals were analysed for evidence of pulsations with none being found. By combining eclipse timings from the four fitted TESS half-sectors with archival eclipse timing data we defined a new high-precision orbital ephemeris.

The properties of both stars were found to be consistent with \textsc{parsec} models for a solar metallicity and an age of 1.7~Gyr. The evolutionary tracks show the stars to be in the second half of their main-sequence lifetime with the more massive star A having evolved farther from the ZAMS.  With two stars of well-constrained properties and age this system lends itself to the calibration of future stellar models, a role which could be further enhanced by the analysis of follow-up spectroscopy to better constrain their atmospheric characteristics.


\section*{Acknowledgements}

We gratefully acknowledge the financial support of the Science and Technologies Facilities Council (STFC) in the form of a PhD studentship. This paper includes data collected by the TESS\ mission and obtained from the MAST data archive at the Space Telescope Science Institute (STScI). Funding for the TESS\ mission is provided by NASA's Science Mission Directorate. STScI is operated by the Association of Universities for Research in Astronomy, Inc., under NASA contract NAS 5–26555.  The following resources were used in the course of this work: the NASA Astrophysics Data System; the SIMBAD database operated at CDS, Strasbourg, France; and the ar$\chi$iv scientific paper preprint service operated by Cornell University. Finally, we thank Jerzy Kreiner and Bartek Zakrzewski for providing TIDAK ephemeris data for \targ\ via private communication. 



\begin{thebibliography}{10}
\newcommand{\enquote}[1]{`(#1)'}

\bibitem{Andersen91aarv}
J.~{Andersen}, \textit{A\&ARv}, \textbf{3}, 91, 1991.

\bibitem{Torres++10aarv}
G.~{Torres}, J.~{Andersen} \& A.~{Gim{\'e}nez}, \textit{A\&ARv}, \textbf{18},
  67, 2010.

\bibitem{Me15aspc}
J.~{Southworth}, in \textit{Living Together: Planets, Host Stars and Binaries}
  (S.~M. {Rucinski}, G.~{Torres} \& M.~{Zejda}, eds.), 2015,
  \textit{Astronomical Society of the Pacific Conference Series}, vol. 496, p.
  321.

\bibitem{ClaretTorres16aa}
A.~{Claret} \& G.~{Torres}, \textit{\aap}, \textbf{592}, A15, 2016.

\bibitem{Tkachenko+20aa}
A.~{Tkachenko} \textit{et~al.}, \textit{\aap}, \textbf{637}, A60, 2020.

\bibitem{Me21univ}
J.~{Southworth}, \textit{Universe}, \textbf{7}, 369, 2021.

\bibitem{Deleuil+18aa}
M.~{Deleuil} \textit{et~al.}, \textit{\aap}, \textbf{619}, A97, 2018.

\bibitem{Borucki16rpp}
W.~J. {Borucki}, \textit{Reports on Progress in Physics}, \textbf{79}, 036901,
  2016.

\bibitem{Howell+14pasp}
S.~B. {Howell} \textit{et~al.}, \textit{\pasp}, \textbf{126}, 398, 2014.

\bibitem{Ricker+15jatis}
G.~R. {Ricker} \textit{et~al.}, \textit{Journal of Astronomical Telescopes,
  Instruments, and Systems}, \textbf{1}, 014003, 2015.

\bibitem{Strohmeier+67ibvs}
W.~Strohmeier \& H.~Ott, \textit{IBVS}, \textbf{195}, 1, 1967.

\bibitem{Popper83aj}
D.~M. Popper, \textit{AJ}, \textbf{88}, 1242, 1983.

\bibitem{Kukarkin+72ibvs}
B.~V. Kukarkin \textit{et~al.}, \textit{IBVS}, \textbf{717}, 1, 1972.

\bibitem{Chen75acta}
K.~Y. Chen, \textit{Acta Astron}, \textbf{25}, 89, 1975.

\bibitem{MauderAmmann76}
H.~Mauder \& M.~Ammann, \textit{Mitteilungen der Astronomischen Gesellschaft
  Hamburg}, \textbf{38}, 231, 1976.

\bibitem{PopperDumont77aj}
D.~M. Popper \& P.~J. Dumont, \textit{AJ}, \textbf{82}, 216, 1977.

\bibitem{BrancewiczDworak80acta}
H.~K. Brancewicz \& T.~Z. Dworak, \textit{Acta Astron}, \textbf{30}, 501, 1980.

\bibitem{Popper80araa}
D.~M. Popper, \textit{ARA\&A}, \textbf{18}, 115, 1980.

\bibitem{WolfKern83apjs}
G.~W. Wolf \& J.~T. Kern, \textit{ApJS}, \textbf{52}, 429, 1983.

\bibitem{PerryChristodoulou96pasp}
C.~L. Perry \& D.~M. Christodoulou, \textit{PASP}, \textbf{108}, 772, 1996.

\bibitem{Nordstrom+97aas}
B.~Nordstr{\"o}m \textit{et~al.}, \textit{A\&AS}, \textbf{126}, 21, 1997.

\bibitem{Gaia21aa}
{Gaia Collaboration} \textit{et~al.}, \textit{\aap}, \textbf{649}, A1, 2021.

\bibitem{CannonPickering20anhar}
A.~J. {Cannon} \& E.~C. {Pickering}, \textit{Annals of Harvard College
  Observatory}, \textbf{95}, 1, 1920.

\bibitem{Hipparcos97}
ESA (ed.), \textit{{The HIPPARCOS and TYCHO catalogues. Astrometric and
  photometric star catalogues derived from the ESA HIPPARCOS Space Astrometry
  Mission}}, \textit{ESA Special Publication}, vol. 1200, 1997.

\bibitem{Hog+00aa}
E.~{H{\o}g} \textit{et~al.}, \textit{\aap}, \textbf{355}, L27, 2000.

\bibitem{Stassun+19aj}
K.~G. {Stassun} \textit{et~al.}, \textit{\aj}, \textbf{158}, 138, 2019.

\bibitem{Cutri+03book}
R.~M. {Cutri} \textit{et~al.}, \textit{{2MASS All Sky Catalog of point
  sources.}} (The IRSA 2MASS All-Sky Point Source Catalog, NASA/IPAC Infrared
  Science Archive., Caltech, US), 2003.

\bibitem{HoukSmith-Moore88}
N.~{Houk} \& M.~{Smith-Moore}, \textit{{Michigan Catalogue of Two-dimensional
  Spectral Types for the HD Stars. Volume 4}} (University of Michigan), 1988.

\bibitem{Lightkurve+18}
{Lightkurve Collaboration} \textit{et~al.}, \enquote{Lightkurve: {{Kepler}} and
  {{TESS}} Time Series Analysis in {{Python}}}, Astrophysics Source Code
  Library, 2018.

\bibitem{Astropy22apj}
{Astropy Collaboration} \textit{et~al.}, \textit{ApJ}, \textbf{935}, 167, 2022.

\bibitem{Jenkins+16spie}
J.~M. {Jenkins} \textit{et~al.}, in \textit{Software and Cyberinfrastructure
  for Astronomy IV} (G.~{Chiozzi} \& J.~C. {Guzman}, eds.), 2016,
  \textit{Society of Photo-Optical Instrumentation Engineers (SPIE) Conference
  Series}, vol. 9913, p. 99133E.

\bibitem{Me++04mn2}
J.~{Southworth}, P.~F.~L. {Maxted} \& B.~{Smalley}, \textit{\mnras},
  \textbf{351}, 1277, 2004.

\bibitem{Me13aa}
J.~{Southworth}, \textit{\aap}, \textbf{557}, A119, 2013.

\bibitem{ClaretSouthworth22aa}
A.~Claret \& J.~Southworth, \textit{A\&A}, \textbf{664}, A128, 2022.

\bibitem{UytterhoevenMoya+11a&a}
K.~Uytterhoeven \textit{et~al.}, \textit{A\&A}, \textbf{534}, A125, 2011.

\bibitem{Me08mn}
J.~{Southworth}, \textit{\mnras}, \textbf{386}, 1644, 2008.

\bibitem{Kreiner04acta}
J.~M. Kreiner, \textit{Acta Astron}, \textbf{54}, 207, 2004.

\bibitem{Me21obs5}
J.~{Southworth}, \textit{The Observatory}, \textbf{141}, 234, 2021.

\bibitem{Duflot+95aas}
M.~Duflot, P.~Figon \& N.~Meyssonnier, \textit{A\&AS}, \textbf{114}, 269, 1995.

\bibitem{Gontcharov06al}
G.~A. Gontcharov, \textit{AL}, \textbf{32}, 759, 2006.

\bibitem{Maxted+20mnras}
P.~F.~L. Maxted \textit{et~al.}, \textit{MNRAS}, \textbf{498}, 332, 2020.

\bibitem{Me++05aa}
J.~{Southworth}, P.~F.~L. {Maxted} \& B.~{Smalley}, \textit{\aap},
  \textbf{429}, 645, 2005.

\bibitem{Hilditch01book}
R.~W. {Hilditch}, \textit{{An Introduction to Close Binary Stars}} (Cambridge
  University Press, Cambridge, UK), 2001.

\bibitem{Prsa+16aj}
A.~{Pr{\v{s}}a} \textit{et~al.}, \textit{\aj}, \textbf{152}, 41, 2016.

\bibitem{Girardi+2002aa}
L.~Girardi \textit{et~al.}, \textit{A\&A}, \textbf{391}, 195, 2002.

\bibitem{Kervella+04aa}
P.~Kervella \textit{et~al.}, \textit{A\&A}, \textbf{426}, 297, 2004.

\bibitem{Bressan+12mnras}
A.~Bressan \textit{et~al.}, \textit{MNRAS}, \textbf{427}, 127, 2012.

\bibitem{Dotter16apjs:MIST}
A.~Dotter, \textit{ApJS}, \textbf{222}, 8, 2016.

\bibitem{Choi+16apj:MIST}
J.~Choi \textit{et~al.}, \textit{ApJ}, \textbf{823}, 102, 2016.

\bibitem{Holmberg+07aa}
J.~Holmberg, B.~Nordstr{\"o}m \& J.~Andersen, \textit{A\&A}, \textbf{475}, 519,
  2007.

\bibitem{CasagrandeSchonrich+11aa}
L.~Casagrande \textit{et~al.}, \textit{A\&A}, \textbf{530}, A138, 2011.

\end{thebibliography}

\end{document}